\documentclass[final,1p,times,authoryear]{elsarticle}
\usepackage{multicol}
\usepackage{color}
\usepackage{xspace}
\usepackage{pdfwidgets}
\usepackage{enumerate}

\usepackage{epsfig}   
\usepackage{amsmath}
\usepackage{mathptmx}
\usepackage{txfonts}
\usepackage[T1]{fontenc}
\usepackage{ae,aecompl}
\usepackage{graphicx}	% Including figure files
\usepackage{amssymb}	% Extra maths symbols
\usepackage{scalerel}
\usepackage{longtable}
\usepackage{titlesec}
\usepackage{hyperref}

\newcommand{\be}{\begin{equation}}
\newcommand{\e}{\end{equation}}
\newcommand{\bear}{\begin{eqnarray}}
\newcommand{\ear}{\end{eqnarray}}

\def\aj{AJ}
\def\apj{ApJ}

\def\mnras{MNRAS}

\journal{Astronomy and Computing}

\begin{document}

\begin{frontmatter}



\title{Separating the blue cloud and the red sequence using Otsu's method for image segmentation}


\author{Biswajit Pandey}

%\author{Name Biswajit Pandey \corref{cor1}\fnref{label2}}
\ead{biswap@visva-bharati.ac.in}
%\fntext[label2]{}
%\cortext[cor1]{}
\affiliation{organization={ Department of Physics,
  Visva-Bharati University},%Department and Organization
            addressline={}, 
            city={Santiniketan},
            postcode={731235}, 
            state={West Bengal},
            country={India}}

\begin{abstract}
The observed colour bimodality allows a classification of the galaxies
into two distinct classes: the `blue cloud' and the `red
sequence'. Such classification is often carried out using empirical
cuts in colour and other galaxy properties that lack solid
mathematical justifications. We propose a method for separating the
galaxies in the `blue cloud' and the `red sequence' using Otsu's
thresholding technique for image segmentation. We show that this
technique is insensitive to the choice of binning. It provides a
robust and parameter-free method for the classification of the red and
blue galaxies based on the minimization of the intra-class variance
and maximization of the inter-class variance. We also apply an
iterative triclass thresholding technique based on Otsu's method to
improve the classification. The galaxy colour is known to depend on
the stellar mass and the luminosity of galaxies. We obtain the
dividing lines between the two populations in the colour-stellar mass
plane and the colour-absolute magnitude plane by employing these
methods in a number of independent stellar mass bins and absolute
magnitude bins.

\end{abstract}




\end{frontmatter}


%\section{}
%\label{}





\section{Introduction}

The galaxies are the fundamental unit of the large-scale
structures. They come in various shapes, sizes, luminosity, mass,
colour, star formation rate and metallicity. Classifying the galaxies
based on their physical properties helps us to understand their
formation and evolution.

The colour of a galaxy is defined as the ratio of fluxes in two
different filters. It is considered one of the fundamental properties
of a galaxy that characterizes its stellar population. It is now well
known that the observed distribution of galaxy colour is strongly
bimodal \citep{strateva, blanton03, bell1, balogh, baldry04}. The
observed colour distribution reveals two distinct peaks corresponding
to a `blue cloud' and a `red sequence'. The `blue cloud' predominantly
hosts the active star-forming galaxies with younger stellar
populations, lower stellar mass and disk-like morphology
\citep{strateva, blanton03, kauffmann03, baldry04}. On the other hand,
the galaxies in the `red sequence' have higher stellar mass with an
older stellar population and bulge-dominated morphology.

The bimodal character of the galaxy colour distribution has important
implications for galaxy formation and evolution. The colour bimodality
is known to exist out to $z=1-2$ \citep{bell2,brammer09}. Observations
indicate that the number of massive red galaxies has increased
steadily since $z \sim 1$ \citep{bell2,faber07}. It indicates that the
blue galaxies transform into red ones via the quenching of star
formation. Such quenching may happen through different physical
processes and mechanisms. A sharp decline in star formation rate
between $z=1$ to present \citep{madau96} also hints towards a
significant evolution of the galaxy properties. Such evolution may
have played a decisive role in shaping the observed colour bimodality
in the present Universe. The colour bimodality also depends on
luminosity, stellar mass and environment \citep{balogh, baldry06,
  pandey20}. Any successful model of galaxy formation must be able to
reproduce the observed colour bimodality. The semi-analytic models of
galaxy formation has been widely used in a number of works to explain
the observed colour bimodality
\citep{menci,driver,cattaneo1,cattaneo2,cameron,trayford,nelson,correa19}.

The red and blue galaxies are known to have different two-point
correlation function \citep{zehavi11}, three-point correlation
function \citep{kayo}, genus \citep{hoyle1}, filamentarity
\citep{pandey06}, local dimension \citep{pandey20} and mass function
\citep{drory, taylor15}. These measurements provide important inputs
to the theories of galaxy formation and evolution. One requires an
operational definition of the two classes of galaxies in all such
studies. Separating the blue and red galaxies is not a trivial task,
as no galaxies can be regarded as either truly `blue' or `red' based
on their colours. The primary motivation for such a classification
lies in the observed bimodality of the colour distribution. The
galaxies in the `blue cloud' and the `red sequence' are usually
separated using specific cuts based on empirical arguments
\citep{strateva, baldry04, williams09, bamford09, arnouts13,
  fritz14}. For instance, \citet{strateva} proposed that the red and
blue galaxies can be optimally separated using a colour cut of
$(u-r)=2.22$. \citet{pandey} propose a fuzzy set theory-based method
for classifying the red, blue and transition valley
galaxies. Nevertheless, this method also has some arbitrariness in
selecting the membership function and the associated parameters. A
number of works have been carried out to distinguish the red and blue
galaxy populations in the GAMA survey. \citet{taylor15} use the
phenomenology of the colour–mass diagrams to provide objective
operational definitions for the red and blue galaxies in the GAMA
survey. \citet{bremer18} divide the galaxies into three broad colour
bins corresponding to red, green and blue galaxies based on the
surface density of points in the colour-mass plane. \citet{turner19}
use k-means unsupervised clustering and \citet{holwerda22} use
Self-Organizing Maps, an unsupervised machine learning technique to
segregate the galaxies according to their properties. Ideally, it
would be most desirable to have a method that can divide the two
populations using some mathematically justified definition. One such
method for statistical decision-making is Otsu's thresholding
technique \citep{otsu79}, originally proposed by Nobuyuki Otsu for
image segmentation. It is a parameter-free method for separating the
foreground pixels from the background. Over the years, it has found
many important applications in remote sensing, robotic mapping and
navigation and identifying tumors. The Otsu's method has been also
applied for object detection in astronomical images \citep{zheng15,
  gong23}.

In this work, we propose an automated method for separating the
galaxies in the `blue cloud' and the `red sequence' for any given data
set using Otsu's algorithm \citep{otsu79} for image segmentation. We
also implement an improved iterative triclass thresholding technique
\citep{cai14} based on Otsu's method to classify the red and blue
galaxies. It provides a nearly parameter-free method for classifying
the red and blue galaxies.

\section{SDSS data}
We apply the proposed method to the data from the Sloan Digital Sky
Survey (SDSS) \citep{york} which is the largest galaxy survey to
date. It has collected the photometric and spectroscopic information
of more than one million galaxies and quasars in five wave bands
across one-quarter of the entire sky. We use the data from the SDSS
DR16 \citep{ahumada} for the current work. We download the data from
the SDSS SkyServer \footnote{https://skyserver.sdss.org/casjobs/}
using SQL. We identify a contiguous region between the right ascension
$135^{\circ} \leq \alpha \leq 225^{\circ}$ and the declination
$0^{\circ} \leq \delta \leq 60^{\circ}$ and extract the spectroscopic
information of all the galaxies with r-band Petrosian magnitude $ 13.5
\le r_{p} < 17.77$ within redshift $z < 0.3$. These cuts provide us
with a total $376495$ galaxies. We then prepare a volume limited
sample by applying a cut to the K-corrected and extinction corrected
$r$-band absolute magnitude. We apply an absolute magnitude cut of
$-21 \ge M_r \ge -23 $ that corresponds redshift limit $0.041 \le z
\le 0.120$. We finally have a total $103984$ galaxies in our volume
limited sample. We download the stellar mass of these galaxies from a
catalogue based on the Flexible Stellar Population Synthesis model
\citep{conroy09}. A $\Lambda$CDM cosmological model with
$\Omega_{m0}=0.315$, $\Omega_{\Lambda0}=0.685$ and $h=0.674$
\citep{planck18} is used for our analysis. We used the same data set
earlier to classify the red, green and blue galaxies using a fuzzy set
theory based method \citep{pandey}.

\section{Method of Analysis}

\begin{figure*}
\centering
\resizebox{8.0cm}{!}{\rotatebox{0}{\includegraphics{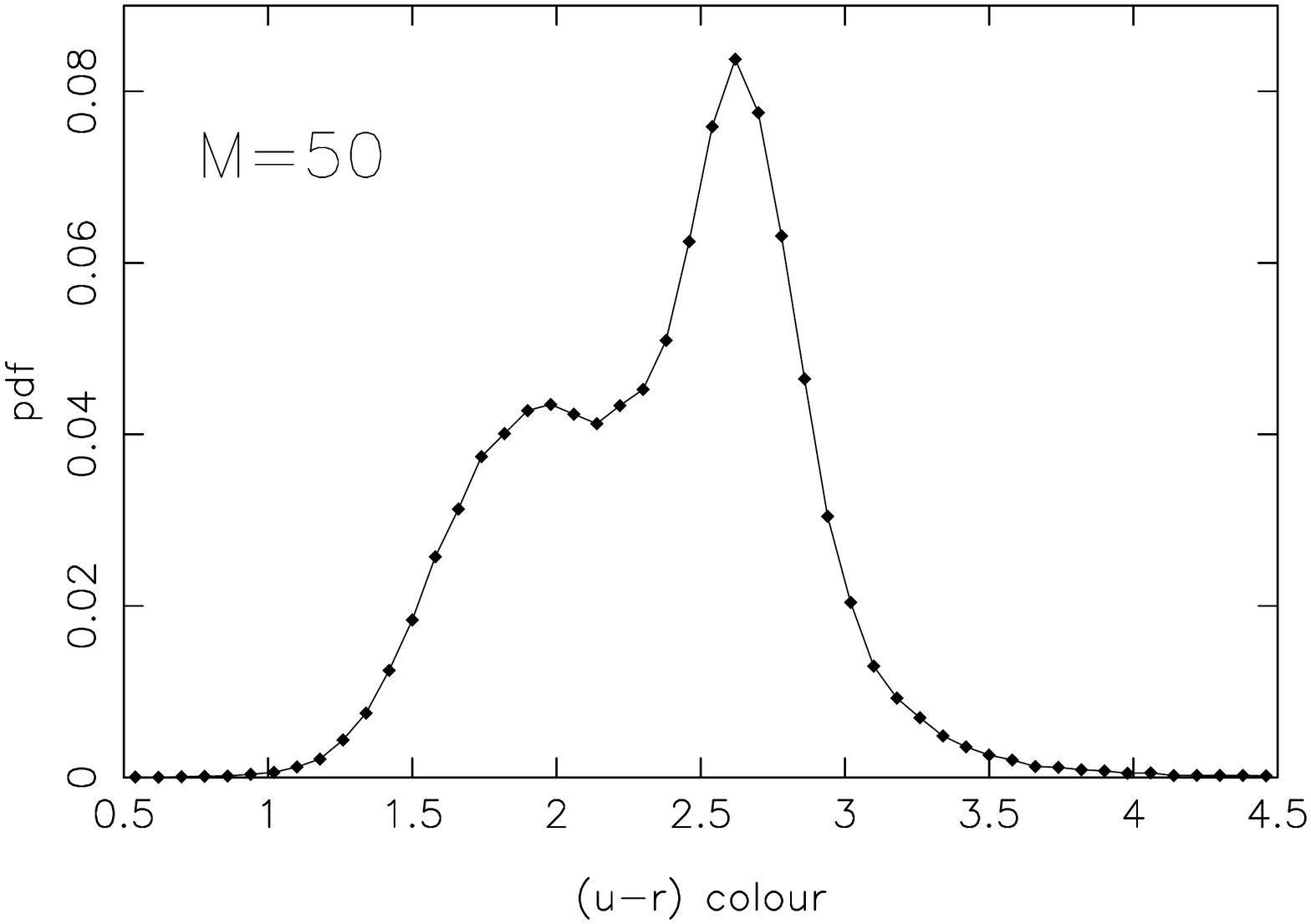}}}
\caption{This shows the $(u-r)$ colour distribution of the SDSS
  galaxies in our volume limited sample. }
\label{fig:pdf}
\end{figure*}

\begin{figure*}
\resizebox{6.65cm}{!}{\rotatebox{0}{\includegraphics{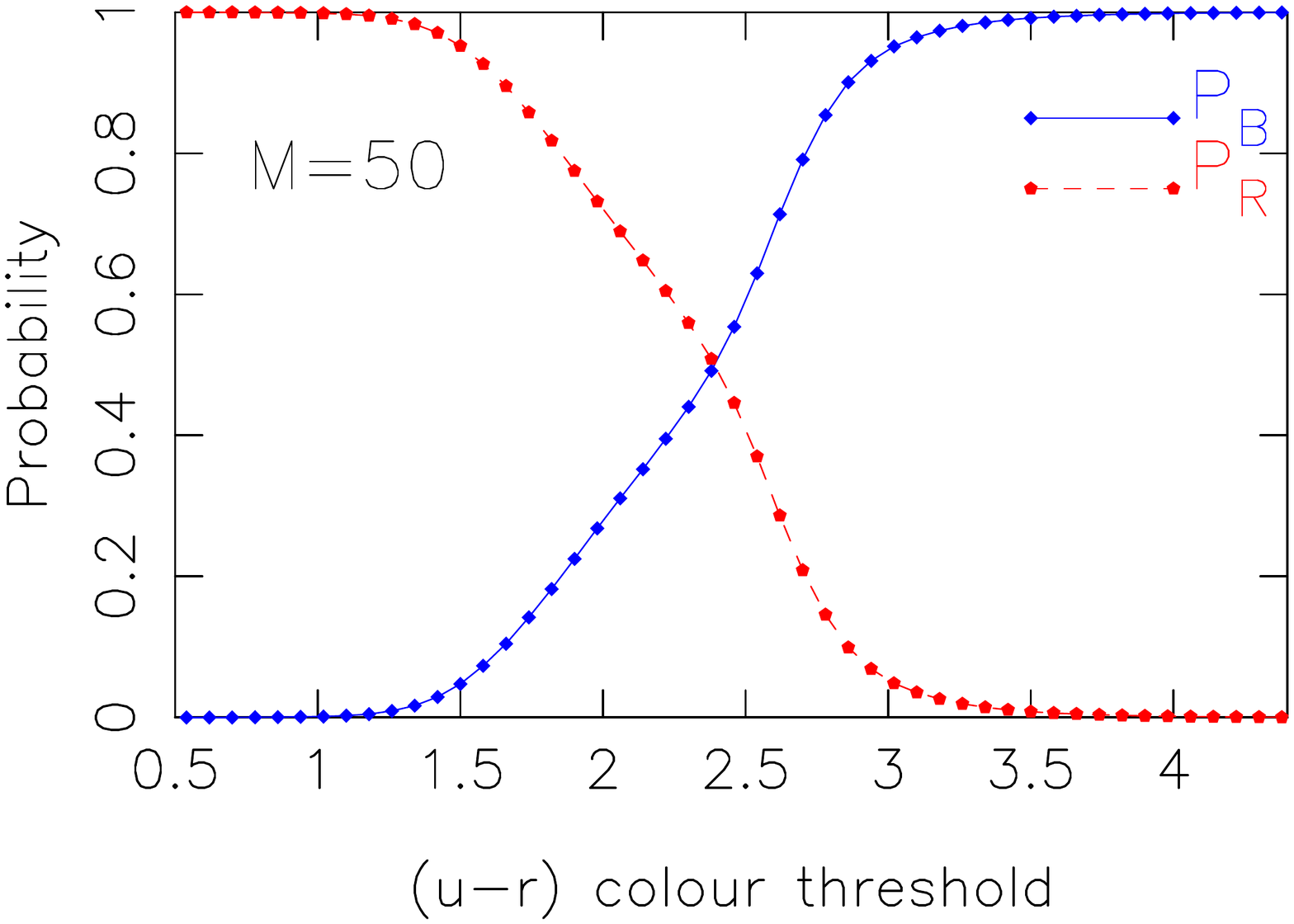}}}%
\resizebox{6.65cm}{!}{\rotatebox{0}{\includegraphics{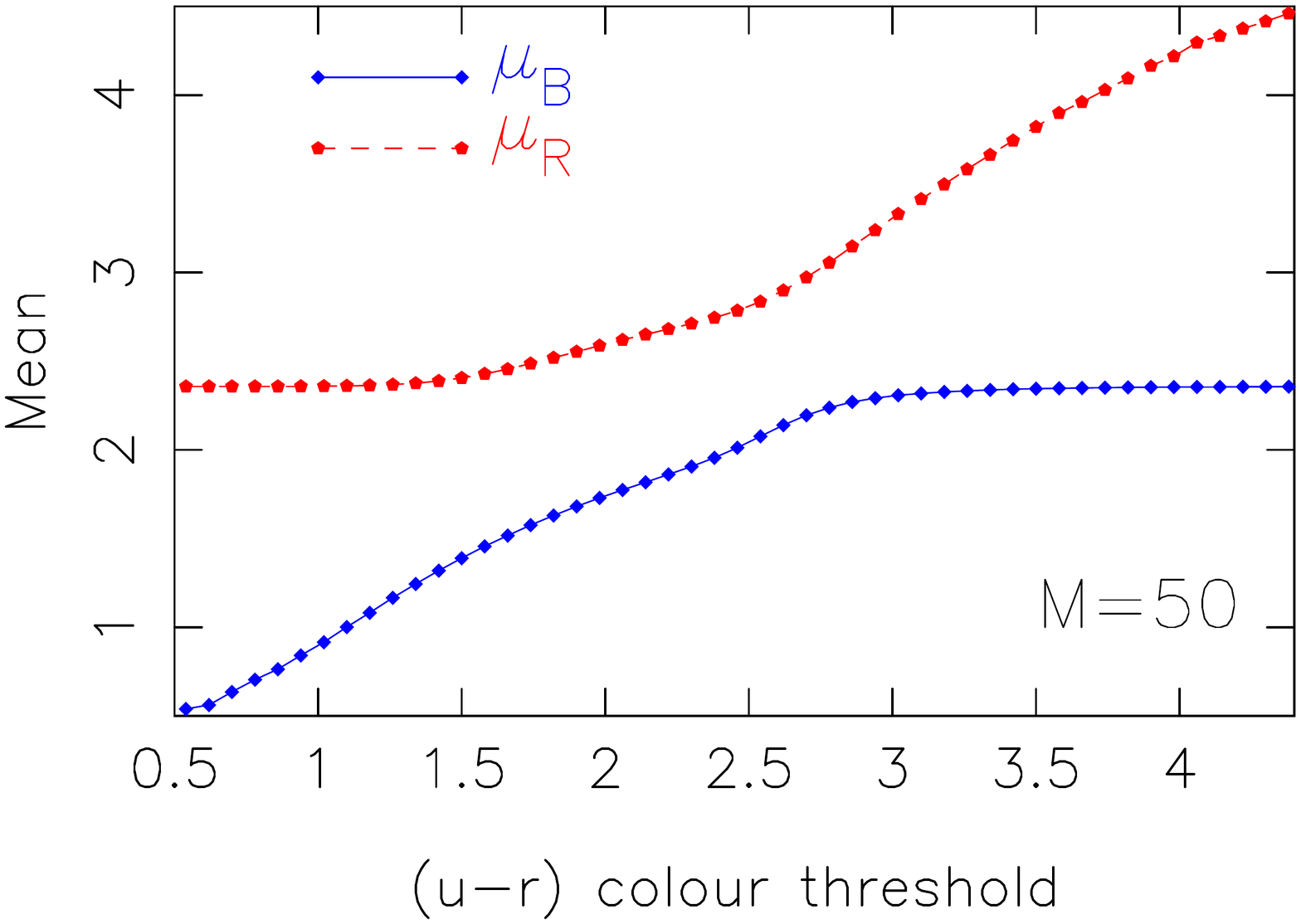}}}\\
\resizebox{6.65cm}{!}{\rotatebox{0}{\includegraphics{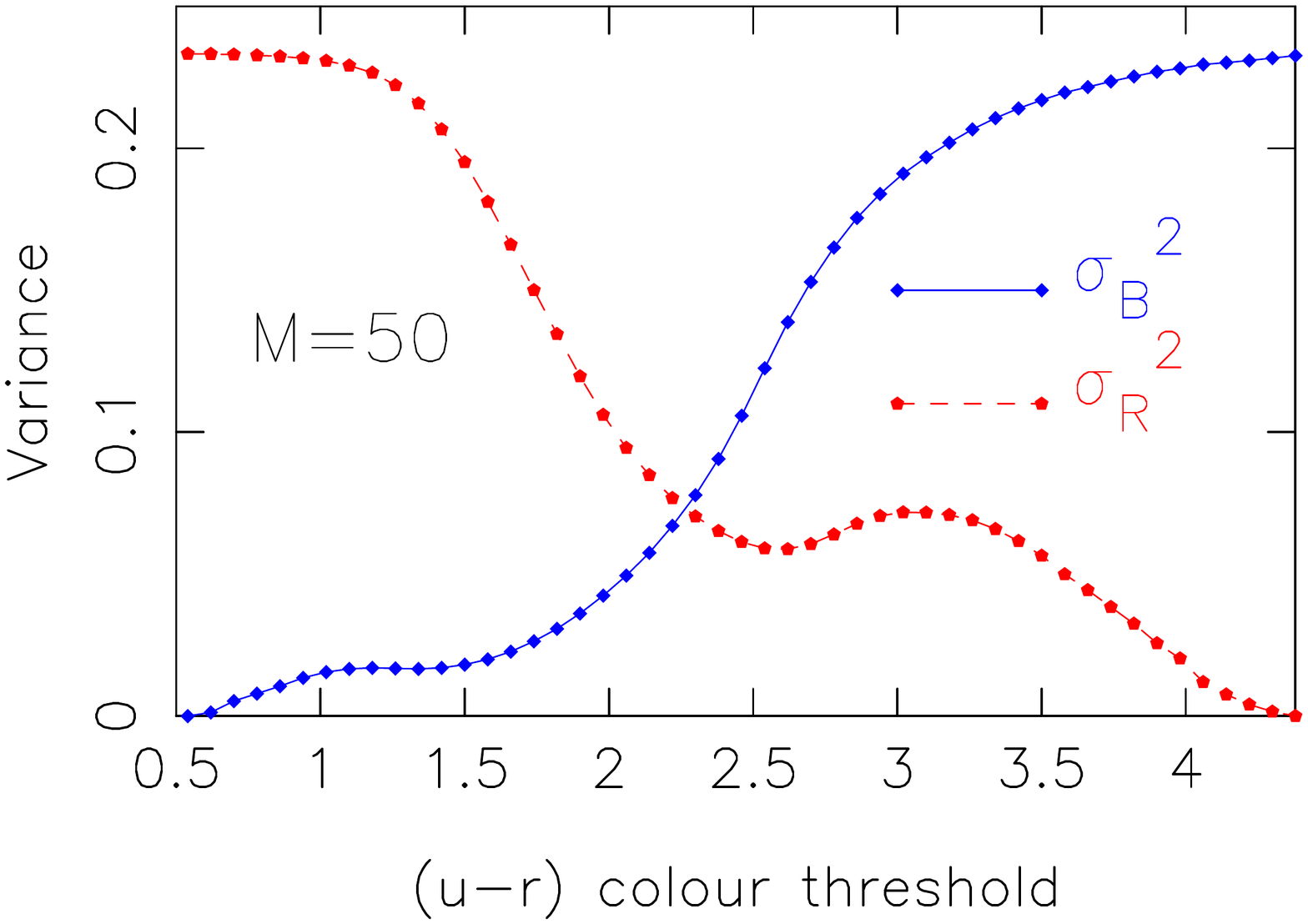}}}%
\resizebox{6.65cm}{!}{\rotatebox{0}{\includegraphics{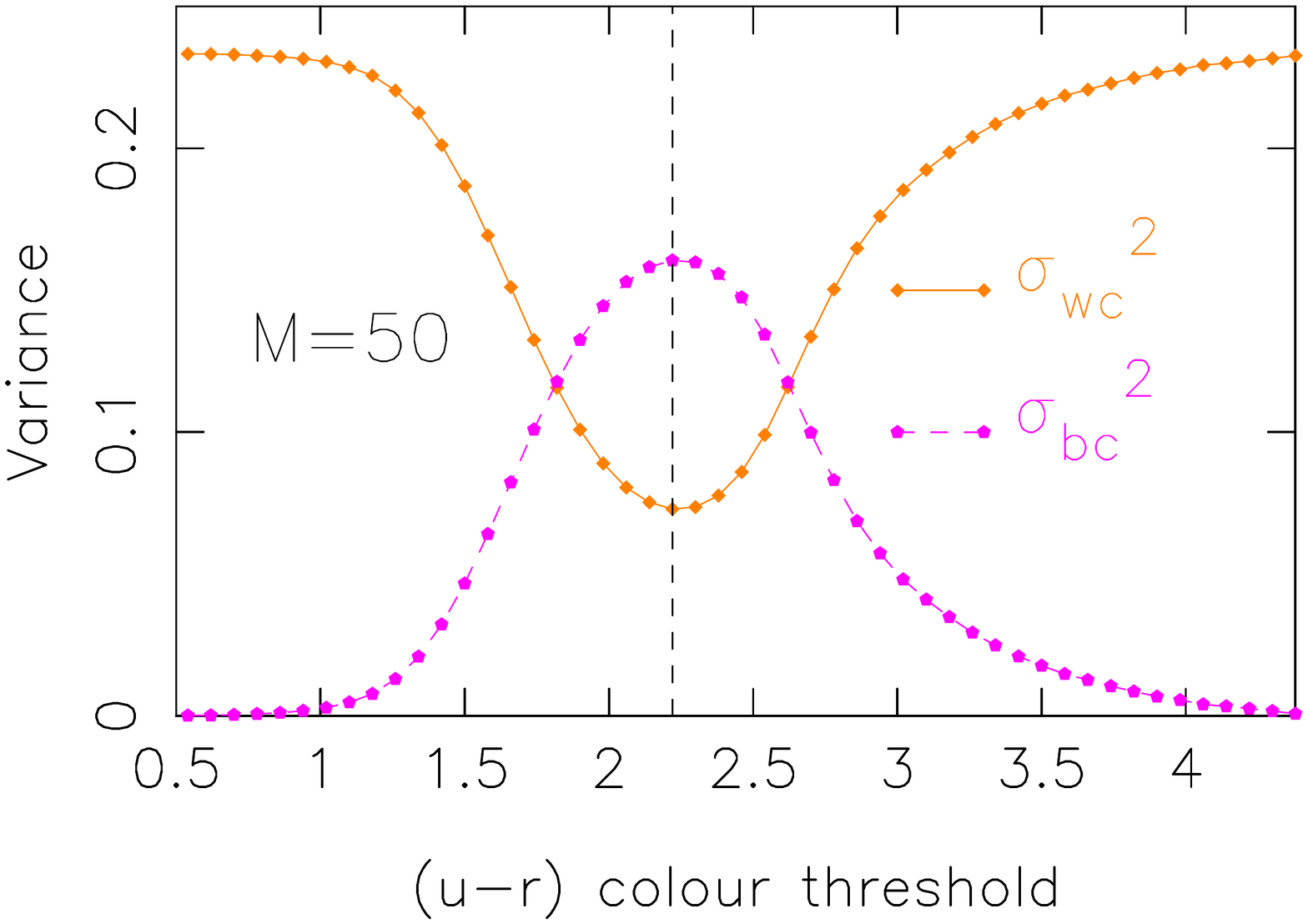}}}\\
\caption{The top left panel shows $P_{B}$ and $P_{R}$ and the top
  right panel shows $\mu_{B}$ and $\mu_{R}$ as a function of the
  $(u-r)$ colour threshold. The variances $\sigma_{B}^2$,
  $\sigma_{R}^2$ and $\sigma_{wc}^2$, $\sigma_{bc}^2$ are shown as a
  function of the $(u-r)$ colour threshold in the bottom left and
  right panels respectively.}
\label{fig:otsu}
\end{figure*}

\begin{figure*}
\resizebox{6.65cm}{!}{\rotatebox{0}{\includegraphics{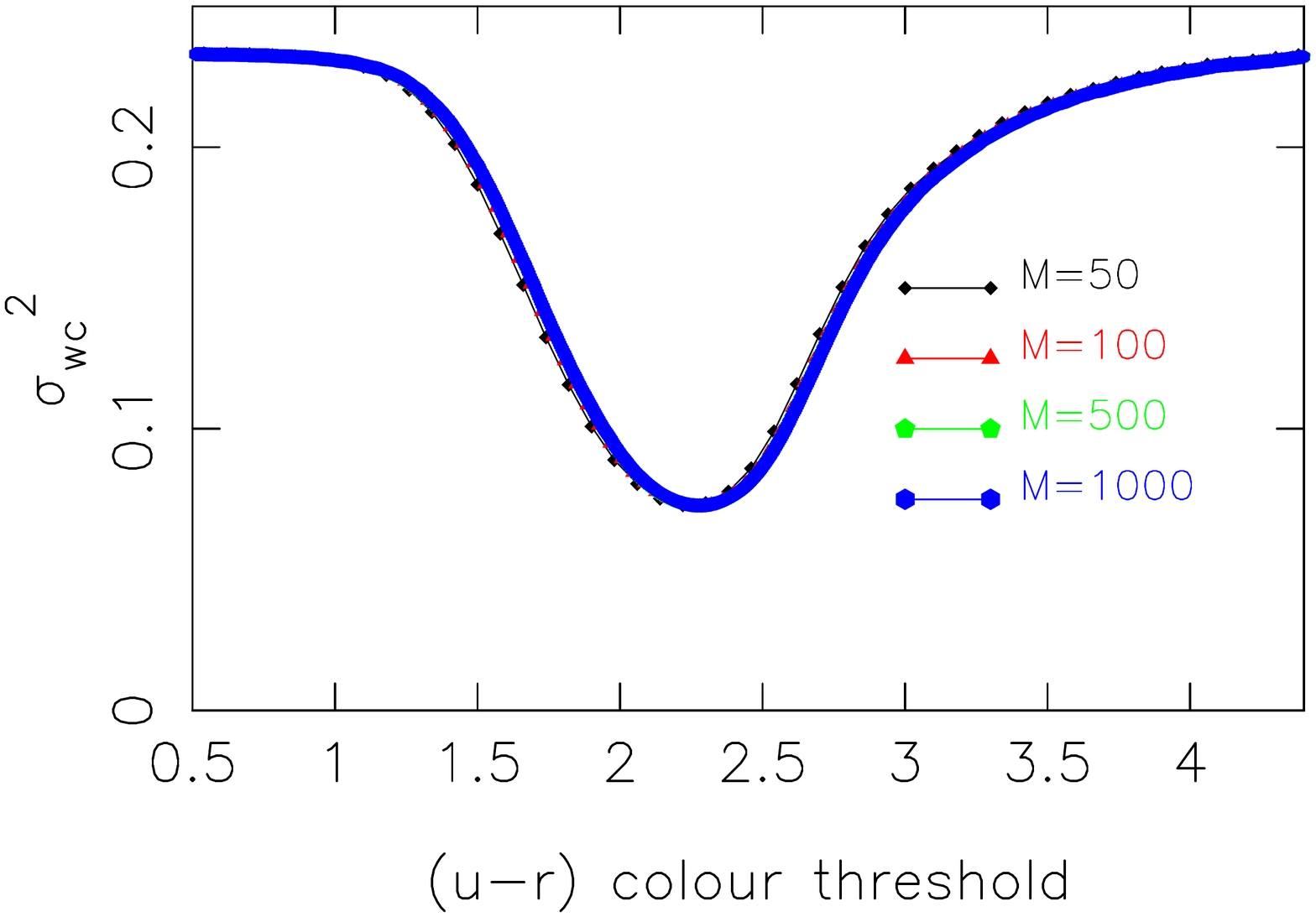}}}%
\resizebox{6.65cm}{!}{\rotatebox{0}{\includegraphics{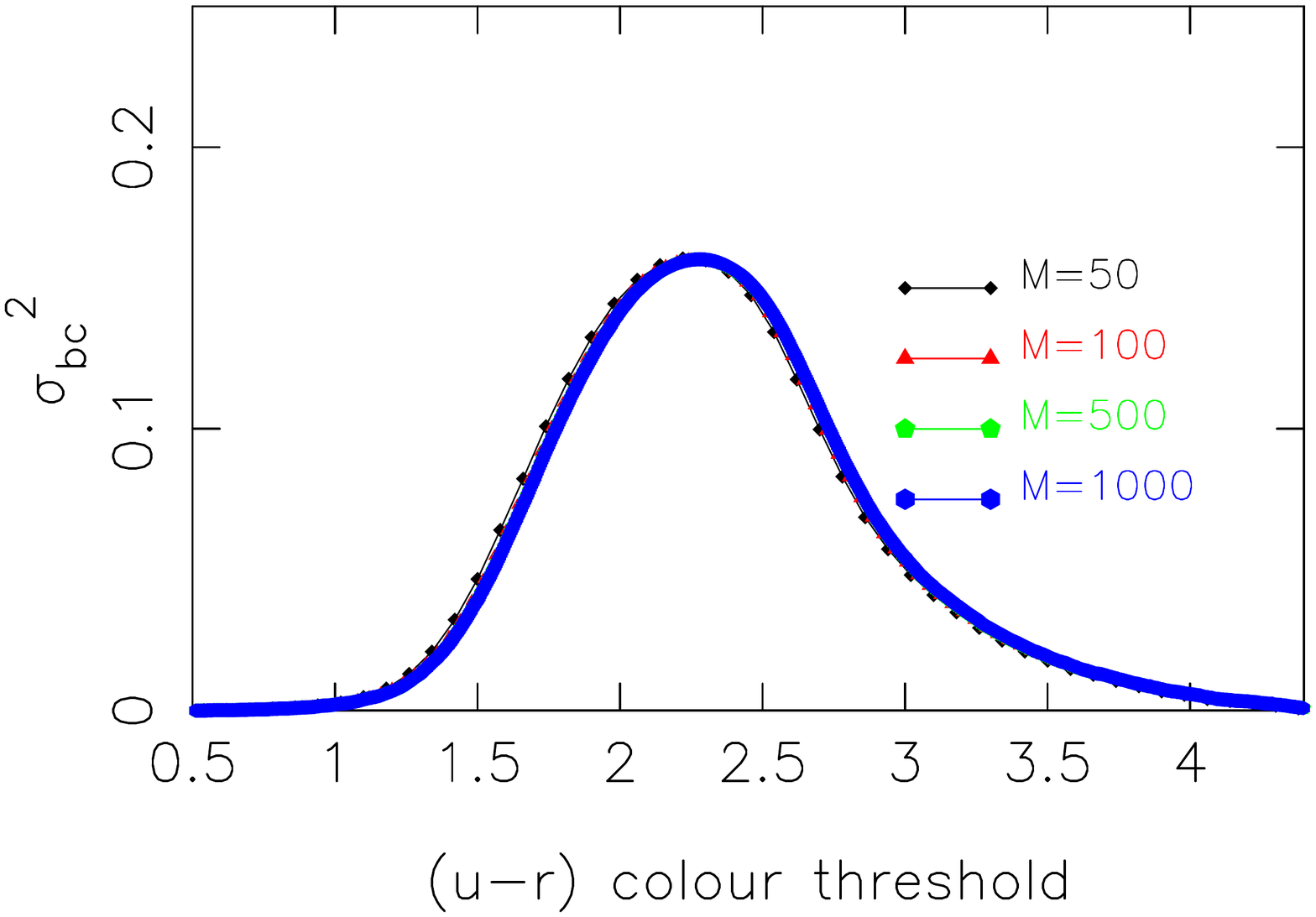}}}\\
\caption{The left panel and the right panel of this figure
  respectively shows the within-class variance $\sigma_{wc}^2$ and the
  between-class variance $\sigma_{bc}^2$ as a function of the $(u-r)$
  colour threshold for different choices of the number of bins.}
\label{fig:nbins}
\end{figure*}

\subsection{Implementing Otsu's thresholding technique to separate the galaxies in the blue cloud and the red sequence}

Otsu's thresholding technique \citep{otsu79} was originally proposed
to separate the foreground pixels from the background pixels in a
grayscale image based on their intensities. The pixels with
intensities greater than a threshold value are marked as foreground,
whereas all pixels with intensity less than or equal to the threshold
are labelled as background. It converts the input grayscale image into
a binary image. The method iterates through all the possible threshold
values and measures the spread of the background and foreground pixels
for each threshold. The method aims to find the threshold value for
which the `intra-class variance' of the two separate pixel populations
is minimum. Interestingly, this maximises the `inter-class variance'
of the foreground and background pixels. An optimal threshold that
ensures either of these will provide the best separation of the two
classes. The method is most efficient when the histogram of the pixel
intensities shows a bimodal nature.

The $(u-r)$ colour distribution of the galaxies in the SDSS is
strongly bimodal \citep{strateva, balogh, baldry04}
(\autoref{fig:pdf}). It motivates us to use Otsu's thresholding
technique for optimally separating the galaxies in the blue cloud and
the red sequence.

We adopt Otsu's method for this work and outline the primary steps
involved in this classification. We consider all the galaxies in our
volume limited sample and calculate the histogram of the $(u-r)$
colour using a specific number of bins. We normalize the histogram of
the $(u-r)$ colour by the total number of galaxies $N=\sum_{i=1}^{M}
n_i$ where $n_i$ is the number of galaxies in the $i^{th}$ colour bin
and $M$ is the total number of bins used in the analysis. This will
ensure $\sum_{i=1}^{M} p_{i}=1$ where $p_{i}=\frac{n_{i}}{N}$. The
resulting probability distribution can be used to calculate the
probabilities of class occurrence for the blue cloud and the red
sequence. Let us denote the blue cloud and the reds sequence with $B$
and $R$ respectively. If the threshold corresponds to the $k^{th}$ bin
then the bins $[1,....,k]$ belongs to the class $B$ and the class $R$
is represented by the bins $[k+1,....,M]$. The probabilities of the
class occurrences for the class $B$ and $R$ can be respectively
written as,\\

\begin{eqnarray}
  P_{B}=\sum_{i=1}^{k} p_{i}=w(k)
\label{eq:weights1}
\end{eqnarray}
and
\begin{eqnarray}
  P_{R}=\sum_{i=k+1}^{M} p_{i}=1-w(k).
\label{eq:weights2}
\end{eqnarray}

We can use these to calculate the class means for each threshold. They are respectively given by,\\

\begin{eqnarray}
  \mu_{B}=\frac{\sum_{i=1}^{k} c_{i} p_{i}}{P_{B}}=\frac{\mu_{k}}{w(k)}
\label{eq:mu1}
\end{eqnarray}
and
\begin{eqnarray}
  \mu_{R}=\frac{\sum_{i=k+1}^{M} c_{i} p_{i}}{P_{R}}=\frac{\mu_{T}-\mu_{k}}{1-w(k)}
\label{eq:mu2}
\end{eqnarray}
where, $c_{i}$ is the $(u-r)$ colour corresponding to the $i^{th}$
bin, $\mu_{k}=\sum_{i=1}^{k} c_{i} p_{i}$ is the mean upto the
$k^{th}$ bin and $\mu_{T}=\sum_{i=1}^{M} c_{i} p_{i}$ is the mean of
the entire distribution. It may be noted that $P_{B}+P_{R}=1$ and
$\mu_{T}=P_{B}\,\mu_{B}+P_{R}\,\mu_{R}$ for each and every threshold.

Similarly, we can also estimate the class variances as,

\begin{eqnarray}
  \sigma_{B}^{2}=\frac{\sum_{i=1}^{k} (c_{i}-\mu_{B})^2 p_{i}}{P_{B}}
\label{eq:sigma1}
\end{eqnarray}
and
\begin{eqnarray}
  \sigma_{R}^{2}=\frac{\sum_{i=k+1}^{M} (c_{i}-\mu_{R})^2 p_{i}}{P_{R}}
\label{eq:sigma2}
\end{eqnarray}

We can define the classification threshold in two different ways: (i)
by minimizing the within-class variance or intra-class variance
$\sigma_{wc}^2$ or (ii) by maximizing the between-class variance or
the inter-class variance $\sigma_{bc}^2$.

The within-class variance $\sigma_{wc}^2$ and the between-class
variance $\sigma_{bc}^2$ can be respectively written as,
\begin{eqnarray}
   \sigma_{wc}^2=P_{B}\, \sigma_{B}^{2}+P_{R}\, \sigma_{R}^{2}
\label{eq:intra}
\end{eqnarray}
and
\begin{eqnarray}
   \sigma_{bc}^2=P_{B}\,P_{R}\,(\mu_{B}-\mu_{R})^2
\label{eq:inter}
\end{eqnarray}

The total variance $\sigma_{T}^2$ is the sum of the intra-class and
inter-class variances,
\begin{eqnarray}
\sigma_{T}^2=\sigma_{wc}^2+\sigma_{bc}^2
\label{eq:total}
\end{eqnarray}

It may be noted that both $\sigma_{wc}^2$ and $\sigma_{bc}^2$ depend
on the chosen threshold, but $\sigma_{T}^2$ is independent of the
threshold. Otsu's algorithm iteratively searches for the threshold
that minimizes the intra-class variance or maximizes the inter-class
variance. Fortunately, the threshold that minimizes $\sigma_{wc}^2$
also maximizes $\sigma_{bc}^2$. So one can choose either
\autoref{eq:intra} or \autoref{eq:inter} to determine the optimal
threshold for the classification. $\sigma_{wc}^2$ involves the
second-order statistics whereas $\sigma_{bc}^2$ is only based on the
first-order statistics. Consequently, it is easier to calculate the
desired threshold using $\sigma_{bc}^2$. In this work, we use both
$\sigma_{wc}^2$ and $\sigma_{bc}^2$ to classify the galaxies in the
blue cloud and the red sequence.

\begin{table}
\centering
\begin{tabular}{|c|c|c|c|c|}
\hline
Iteration & $\mu_B$ & $\mu_R$ & threshold & $|\Delta|$ threshold\\
\hline
1 & $1.861672$ & $2.681102$ & $2.22$ &\\
2 & $2.074795$ & $2.508090$ & $2.279581$ & $0.059581$\\
3 & $2.188867$ & $2.412384$ & $2.295775$ & $0.016194$\\
4 & $2.244947$ & $2.358102$ & $2.298391$ & $0.002615$\\
5 & $2.273358$ & $2.330187$ & $2.300393$ & $0.002002$\\
6 & $2.287296$ & $2.316201$ & $2.301204$ & $0.000812$\\
\hline
\end{tabular}
\caption{This table shows the smaller mean ($\mu_B$), the larger mean
  ($\mu_R$), the $(u-r)$ threshold and the absolute difference between
  two consecutive thresholds ($|\Delta$|) at each iterations of the
  iterative triclass thresholding method. Here we set a critical
  $|\Delta|=0.001$ and choose the number of bins to be $50$.}
\label{tri}
\end{table}

\subsection{Iterative triclass thresholding technique: an improved version of Otsu's method}

The primary limitation of Otsu's method is that the class with the
larger variance has a greater influence in determining the
classification threshold. It may provide sub-optimal results when one
of the classes has a considerably larger variance. Several
improvements of the standard Otsu's method have been proposed in the
literature. Here, we have chosen an iterative triclass thresholding
technique proposed by \citet{cai14}.

In this method, we first determine the threshold using the standard
Otsu's algorithm. We then separate the galaxies into three classes
based on the mean of the two classes. We define the galaxies in the
blue cloud as those with an $(u-r)$ colour less than the smaller
mean. The galaxies in the red sequence are defined as those having
their $(u-r)$ colour greater than the larger mean. The intervening
region between the two means are defined as the "to-be-determined."
(TBD) class. In the next iteration, the regions already
classified as blue cloud and red sequence are kept unchanged. The
standard Otsu's method is again applied only to the TBD region to
divide it into three classes similarly. We get a new
threshold after each iteration. The iteration procedure stops when the
difference between two consecutive thresholds is less than a preset
value. The TBD region is divided into two classes instead of three at
the last iteration. Finally, the classified blue cloud is the logical
union of all the regions that are previously identified as the blue
cloud throughout the different iterations. The final region consisting
of the red sequence is determined identically.

\begin{figure*}
\centering
\resizebox{10.0cm}{!}{\rotatebox{270}{\includegraphics{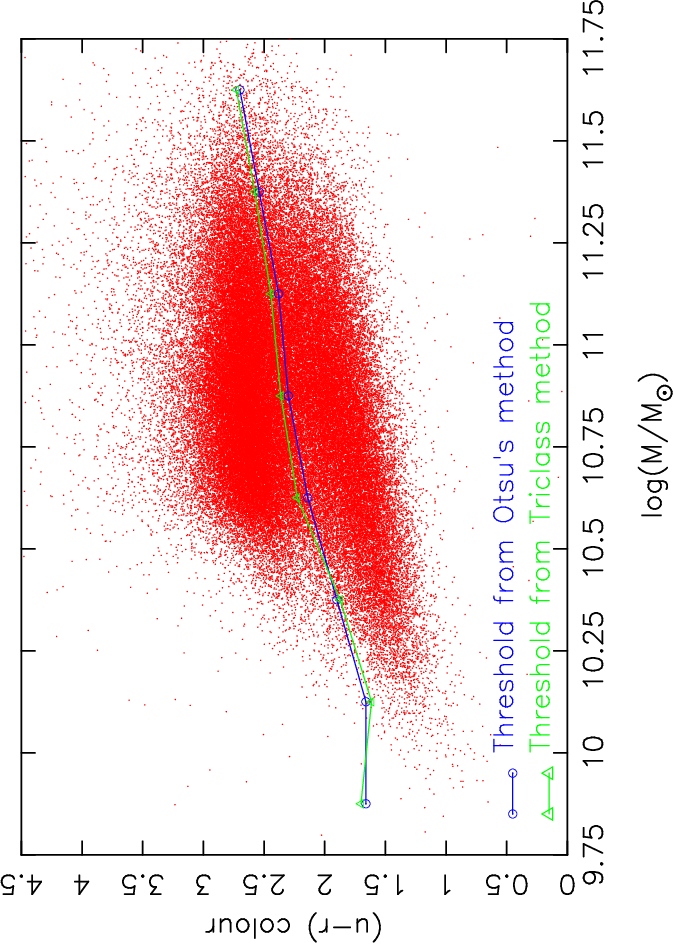}}}
\caption{This shows the $(u-r)$ colour thresholds as a function of
  stellar mass. The dividing lines between the blue cloud and the red
  sequence are shown together for both the Otsu's method and the
  iterative triclass thresholding technique.}
\label{fig:divider1}
\end{figure*}

\begin{figure*}
\centering
\resizebox{10.0cm}{!}{\rotatebox{270}{\includegraphics{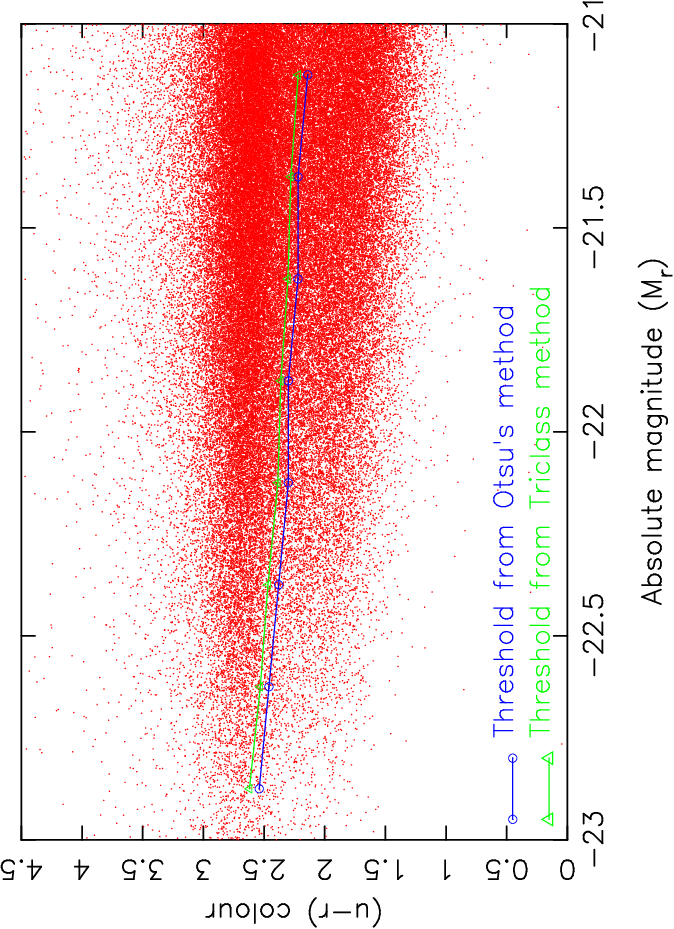}}}
\caption{This shows the $(u-r)$ colour thresholds as a function of
  r-band Absolute magnitude. The boundary lines between the blue cloud
  and the red sequence are shown together for both the Otsu's method
  and the iterative triclass method.}
\label{fig:divider2}
\end{figure*}

\subsection{Results and Conclusions}
We show the probability distribution function of the $(u-r)$ colour
for the SDSS galaxies in \autoref{fig:pdf}. The $(u-r)$ colour
distribution clearly shows a bimodal nature that motivates us to use
the Otsu's thresholding technique to classify the two populations
associated with the two peaks of the distribution.

We show the probabilities of the class occurrences (
\autoref{eq:weights1} and \autoref{eq:weights2}) as a function of the
$(u-r)$ colour threshold for the blue cloud and the red sequence in
the top left panel of \autoref{fig:otsu}. The associated means and
variances (\autoref{eq:mu1}, \autoref{eq:mu2}, \autoref{eq:sigma1} and
\ref{eq:sigma2}) of the two populations as a function of the $(u-r)$
colour threshold are respectively shown in the top right and bottom
left panels of \autoref{fig:otsu}. We compute the within-class
variance $\sigma_{wc}^2$ and the between-class variance
$\sigma_{bc}^2$ of the two populations using \autoref{eq:intra} and
\autoref{eq:inter}. The results are shown in the bottom right panel of
\autoref{fig:otsu}. We find that the within-class variance
$\sigma_{wc}^2$ has its minimum at a $(u-r)$ colour threshold of
$2.22$. We also note that the between-class variance $\sigma_{bc}^2$
has its maximum at $(u-r)=2.22$. Thus $\sigma_{wc}^2$ is minimized and
$\sigma_{bc}^2$ is maximized at exactly the same $(u-r)$ colour
threshold.

It may be noted that the results shown in the \autoref{fig:pdf} and
\autoref{fig:otsu} correspond to a choice of the number of bins
$n=50$. We also test if the locations of the minimum of
$\sigma_{wc}^2$ and the maximum of $\sigma_{bc}^2$ are sensitive to
the choice of the number of bins. We repeat our analysis for $n=100$,
$500$ and $1000$ and show $\sigma_{wc}^2$ and $\sigma_{bc}^2$ as a
function of the $(u-r)$ colour threshold for four different choices of
the number of bins in the left and right panels of
\autoref{fig:nbins}. Interestingly, the locations of the minimum of
$\sigma_{wc}^2$ and the maximum of $\sigma_{bc}^2$ are independent of
the choice of the number of bins. The threshold obtained using this
method is insensitive to the choice of binning and thus provides a
robust technique for classifying the red and blue galaxies.

We propose Otsu's technique as a parameter-free method for the
automated classification of the galaxies in the blue cloud and the red
sequence. It is interesting to note that the $(u-r)$ colour threshold
predicted by Otsu's method in this work matches with the $(u-r)$
colour separator proposed by \citet{strateva}. \citet{strateva}
proposed this as an empirical cut based on the best fit
colour-magnitude relations. We also applied the iterative triclass
thresholding technique based on Otsu's method to improve the
classification. The threshold values obtained after each iteration in
the iterative triclass thresholding scheme are tabulated in Table
\ref{tri}. The iteration is continued until the absolute difference
between two consecutive thresholds ($|\Delta|$) is smaller than a
preset value. We choose $|\Delta|=10^{-3}$ for the present
analysis. It leads to the convergence to an $(u-r)$ colour threshold
of $2.301$ after six iterations. We note that the modified version of
Otsu's method is a nearly parameter-free method that can be used
effectively for separating red and blue galaxies. It may be noted that
the triclass thresholding technique yields a somewhat larger value for
the threshold as compared to the Otsu's method. We know that the
differences in the variances of the two classes may bias the
classification threshold in the Otsu’s method. The triclass method
takes this into account and hence provides an unbiased estimate of the
threshold. Further, there exist clear relationships between colour and
stellar mass or absolute magnitude. Consequently, a single colour
threshold for the galaxies of all masses and luminosity can not be
justified. Hence, we divide the entire sample into a number of
independent stellar mass bins and separately apply the Otsu’s method
and the iterative triclass thresholding technique to each of these
stellar mass bins. We show the dividing lines between the blue cloud
and the red sequence in the colour-stellar mass plane in
\autoref{fig:divider1}. A similar analysis is also carried out using a
number of absolute magnitude bins. The boundary lines between the two
populations in the colour-absolute magnitude plane are shown in
\autoref{fig:divider2}.

The same dataset has been used earlier to divide the red, green and
blue galaxies using a fuzzy set theory based method
\citep{pandey}. This method is also used in an in-depth astronomical
study \citep{das21}. The application of a sharp cut would always
introduce some contaminations due to the presence of the green valley
between the blue cloud and the red sequence. The green valley is
populated by transitioning galaxies that evolve from star-forming to
quiescent systems. One needs to also identify the green valley in
order to separate the blue cloud and the red sequence in an effective
manner. Currently, there are no provision for identifying the green
valley using Otsu’s method or the triclass thresholding technique used
in the present work. \citet{pandey} apply their method to the entire
sample without considering the variations of the colour criteria with
stellar mass or absolute magnitude. However, we consider these
variations in the present work and repeat our analysis in a number of
independent stellar mass bins and absolute magnitude bins. This allows
us to classify the red and blue populations in the colour-stellar mass
plane and colour-absolute magnitude plane. We plan to use the proposed
classification scheme in a future in-depth astronomical study.

A caveat in our analysis is that we use observed colour for
simplicity. The galaxy colours are affected by the reddening due to
redshift and the internal extinction. Ideally, one should use the
rest-frame colours for any such classification \citep{salim14,
  taylor15}.

The distributions of several other galaxy properties, such
as the star formation rate, the stellar mass, the bulge to disk mass
ratio and the stellar age \citep{elbaz07, drory, driver, zibetti17}
also exhibit a bimodal nature. The galaxies in the red sequence are
known to have lower star formation rates and higher stellar masses,
higher stellar ages, higher bulge-to-disk mass ratios as compared to
the galaxies in the blue cloud. Otsu's method can be applied to
classify the galaxies based on each of these properties. One can
employ simultaneous cuts on multiple properties to improve the
red-blue classification proposed in this work.

The primary aim of the present work is to find a mathematically
justified definition of the red and blue populations in a given data
set. We conclude that Otsu's thresholding technique provides us with a
robust and parameter-free method for classifying the galaxies based on
the bimodal distributions of galaxy colour.

\section{Acknowledgementt}
The author thanks the anonymous reviewers for the valuable comments
and suggestions. The author would like to acknowledge the financial
support from the SERB, DST, Government of India through the project
CRG/2019/001110. The author also thanks Suman Sarkar for the help with
the SDSS data. Finally, thanks to the SDSS team for making the data
public.

\end{document}